**Reduced collinearity, low-dimensional cluster expansion model for adsorption of halides (Cl, Br) on Cu(100) surface using principal component analysis**

Bibek Dash[1,2#], Suhail Haque[1#], Abhijit Chatterjee[1]*

[1] Department of Chemical Engineering, Indian Institute of Technology Bombay, Mumbai 400076

[2] Process Engineering & Instrumentation Department, CSIR – Institute of Minerals and Materials Technology, Bhubaneswar

[#] Equal contribution

*Email: abhijit@che.iitb.ac.in

**Abstract**

The cluster expansion model (CEM) provides a powerful computational framework for rapid estimation of configurational properties in disordered systems. However, the traditional CEM construction procedure is still plagued by two fundamental problems: (i) even when only a handful of site cluster types are included in the model, these clusters can be correlated and therefore they cannot independently predict the material property, and (ii) typically few tens-hundreds of datapoints are required for training the model. To address the first problem of collinearity, we apply the principal component analysis method for constructing a CEM. Such an approach is shown to result in a low-dimensional CEM that can be trained using a small DFT dataset. We use the *ab initio* thermodynamic modeling of Cl and Br adsorption on Cu(100) surface as an example to demonstrate these concepts. A key result is that a CEM containing 10 effective cluster interactions build with only 8 DFT energies (note, number of training configurations > number of principal components) is found to be accurate and the thermodynamic behavior obtained is consistent with experiments. This paves the way for construction of high-fidelity CEMs with sparse/limited DFT data.

## 1. Introduction

Adsorption of halogens, such as Cl and Br, on metal surfaces is an important process in heterogeneous catalysis, electronics and semiconductor industry [1–7]. A fundamental understanding of the role of halides is possible only through the construction of thermodynamic models that accurately capture the interactions between the halide adlayer and metal surface. Interaction of metal with halogens can be both beneficial and harmful. For instance, copper chloride ($CuCl_2$) is commercially manufactured by chlorination of copper via direct exposure to $Cl_2$ gas [8]. However, when exposed to humid conditions, metals can undergo corrosion, which are facilitated by halides, which can be undesirable and sometimes even harmful (e.g. fire accidents involving electronic devices) [7,9–13]. Adsorbed halides are known to modify properties of transition metal catalysts due to their strong interaction with metals [5,14–18]. In electrochemical systems, halide on metal surfaces constitute one of the simplest models for specific adsorption and double layer formation. In case of electrochemical reactions, adsorbed halides on Cu are sometimes known to enhance the electrocatalytic activity [19–22].

We study two gas-phase systems, namely, Cl/Cu(100) and Br/Cu(100), that exhibit interesting phase behavior. The adsorption/binding energy per adsorbed atom is defined as

$$E_{ads} = \frac{E_{nX/M} - E_M - nE_X}{n}. \qquad (1)$$

Here $E_{nX/M}$ is the energy of the system comprising of $n$ adsorbed atoms ($X = Cl$ or $Br$), $E_M$ is the energy associated with the bare surface and $E_X$ is the energy of a halogen atom in vacuum.

The strong electronegative character of halogens cause nearby adsorbed atoms to interact laterally [16,23,24]. The adsorbate-adsorbate interactions are typically repulsive in nature. The adsorption energy can be written as

$$E_{ads} = E_{ads}^0 + \frac{1}{n} E_{ads-ads}. \tag{2}$$

$E_{ads}^0$ is the adsorption energy in the dilute limit where adsorbate-adsorbate interactions are absent. $E_{ads}$ typically becomes more positive with increasing coverage. $E_{ads-ads}$ is a many-body adsorbate-adsorbate interaction term, which incorporates the coverage effect. Dividing $E_{ads-ads}$ by $n$ in Equation (2) yields the average interaction per adsorbed atom. $E_{ads-ads}$ is important at high Cl coverage and are known to result in ordered adlayers. Many experimental studies have been devoted to halide adsorption on Cu and its effect on the surface and electronic properties. Various experimental techniques through STM (scanning tunneling microscopy) imaging [5,25,26], LEED (low-energy electron diffraction) intensity analysis [25,27] and angle-resolved photoemission spectra [28,29] of chemisorbed halides on copper surface have been reported. It has been observed that chloride ions specifically adsorbed on the copper surfaces forming an ordered c(2×2) adlayer structure [26,30,31]. Similar results are obtained in UHV experiments [32], XRD [23,33,34] and X-ray photoelectron diffraction [32,34]. The interactions are also known to indirectly affect the kinetics of surface reactions [10,35–37].

An approximate functional form used to describe adsorbate-adsorbate interactions for a given configuration involves the use of a cluster expansion model (CEM)[38–40]. A CEM for Cl and Br adsorption on Cu(100) is not available in literature. Here, we use the form[41,42]

$$E_{ads-ads} = e_b + \sum_c e_c n_c. \tag{3}$$

In Equation (3), $\{c\}$ refers to different site clusters such as singlets, pairs, triplets, and so on, which are included to describe the adsorbate-adsorbate interactions. $n_c$ refers to the number of clusters of type $c$ found in the particular configuration and $e_c$ is the corresponding cluster interaction coefficient. $e_b$ is a bias term, which should ideally be nearly zero. The CEM would have been exact had clusters of all possible sizes been included. However, since the model is truncated at some point, $e_c$ is termed as the effective cluster interactions (ECI). The ECIs are determined using a regression procedure. Equation (3) is a series expansion in which the features (sites and cluster types) are chosen by the user. Systematic strategies are available for feature selection, e.g., cluster selection[43], genetic algorithm[44,45], compressive sensing[46], decision tree[47], and Bayesian inference[48,49]. The general idea is that sites within a cutoff radius[50] are considered, and clusters formed by these sites, such as pairs and triplets, are used in the model. Quadruplets and large-sized clusters are usually excluded. A major problem is that the clusters are correlated. For example, a high number of triplets also implies a high number of pairs. In conventional regression analysis, it is strictly required that the independent variables should be uncorrelated, which is not the case in such a CEM building exercise.

To address the issue of collinearity in the training dataset, we explore the use of principal component regression (PCR)[51] for CEM training. To the best of our understanding, such an approach has not been explored previously. These methods offer an advantage that although the CEM may include a very large number of cluster types, none of the clusters need to be excluded. A low-dimensional space $\{\Lambda_p\}$ is identified, where $p$ denotes a principal component, which captures the maximal variation in the cluster counts. Regression is performed in a low dimensional space

$$E_{ads-ads} = e_b' + \sum_p \epsilon_p \Lambda_p. \qquad (4)$$

where $\Lambda_p$ is of the form

$$\Lambda_p = \sum_c \theta_c (n_c - \bar{n}_c). \tag{5}$$

$\bar{n}_c$ is the mean value of the cluster count. Use of principal components reduces the number of cluster coefficients $\epsilon_p$ that need to be determined. For instance, only one cluster coefficient needs to be fitted if one principal component is found to be adequate. This allows the CEM to be trained with small DFT datasets. Since the principal components form an orthogonal basis set, the resulting $e_c$ are physically meaningful. Combining Equations (4) and (5) provides expression for $e_c$ in Equation (3).

In section 2, we describe the application of PCR for training of CEM. Computational details (density functional theory, CEM and grand canonical Monte Carlo (GCMC)) are also discussed. Results are discussed in Section 3. Finally, conclusions are provided in Section 4.

## 2. Computational details

### *2.1 Density Functional Theory*

Density functional theory (DFT) calculations are performed using the LCAO simulation engine of QuantumATK software package [52]. Norm-conserving pseudopotentials from the PseudoDojo library [53] and ultra basis sets are used. The Perdew-Burke-Ernzerhof (PBE) exchange-correlation functional [54] is employed. Fermi-Dirac smearing of 0.01 eV is applied for electronic occupations. Brillouin zone is sampled with a $(9 \times 9 \times 9)$ Monkhorst-Pack k-point grid [55] for bulk Cu and a 3x3x1 k-point grid for Cu(100) surface slabs. The self-consistent field convergence criterion is set to $10^{-6}$ eV. The Grimme DFT-D2 method was used to include van der Waals interactions [56]. The global scaling factor S6 was set to 0.75 and the damping parameter d was set to 20.0, as recommended for PBE calculations.

The bulk crystal is subjected to geometry optimization with a $(9 \times 9 \times 9)$ k-point grid. The atomic positions and cell parameters were relaxed until the residual forces are less than 0.01 $eV/Å$ and the stress tensor components is less than 0.01 $GPa$. The lattice parameters obtained are $a = b = c =$3.615 Å and $α = β = γ = 90°$ which is in good agreement with experimental lattice data $a = b = c =$3.61 Å [57]. A surface slab model of $3 × 3$ Cu(100) surface was constructed from the optimized bulk crystal cell. The slab contained 4 atomic layers with the bottom two layers being constrained. A vacuum layer of 15 Å was used to avoid interaction between periodic images. Periodic boundary conditions were applied in all directions. The relaxations were carried out until the residual forces were less than 0.05 $eV/Å$. Figure 1 shows the surface slab model of Cu(100) used for DFT calculations.

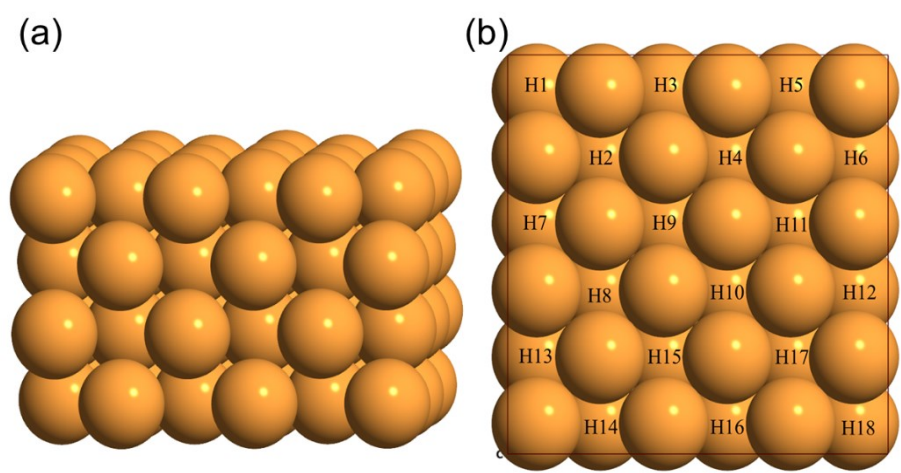

**Figure 1.** Surface slab model of Cu(100); (a) side view; (b) top view showing 18 hollow lattice sites.

Binding of single halide atom on top, bridge and hollow sites on the Cu(100) surface were investigated. Both Cl and Br were found to bound more strongly at the hollow sites (Table S1 in Supporting Information). For this reason, only four-fold hollow sites as represented by H1-H18 in Figure 1(b) are considered. Another key observation is that introduction of a second halide atom at the first nearest neighbor (1NN) position results in significantly weaker binding (Table S2 in Supporting Information). For this reason, such atomic configurations are excluded from our DFT database (see Table S3 in Supporting Information, which provides the site occupation from H1 to H18).

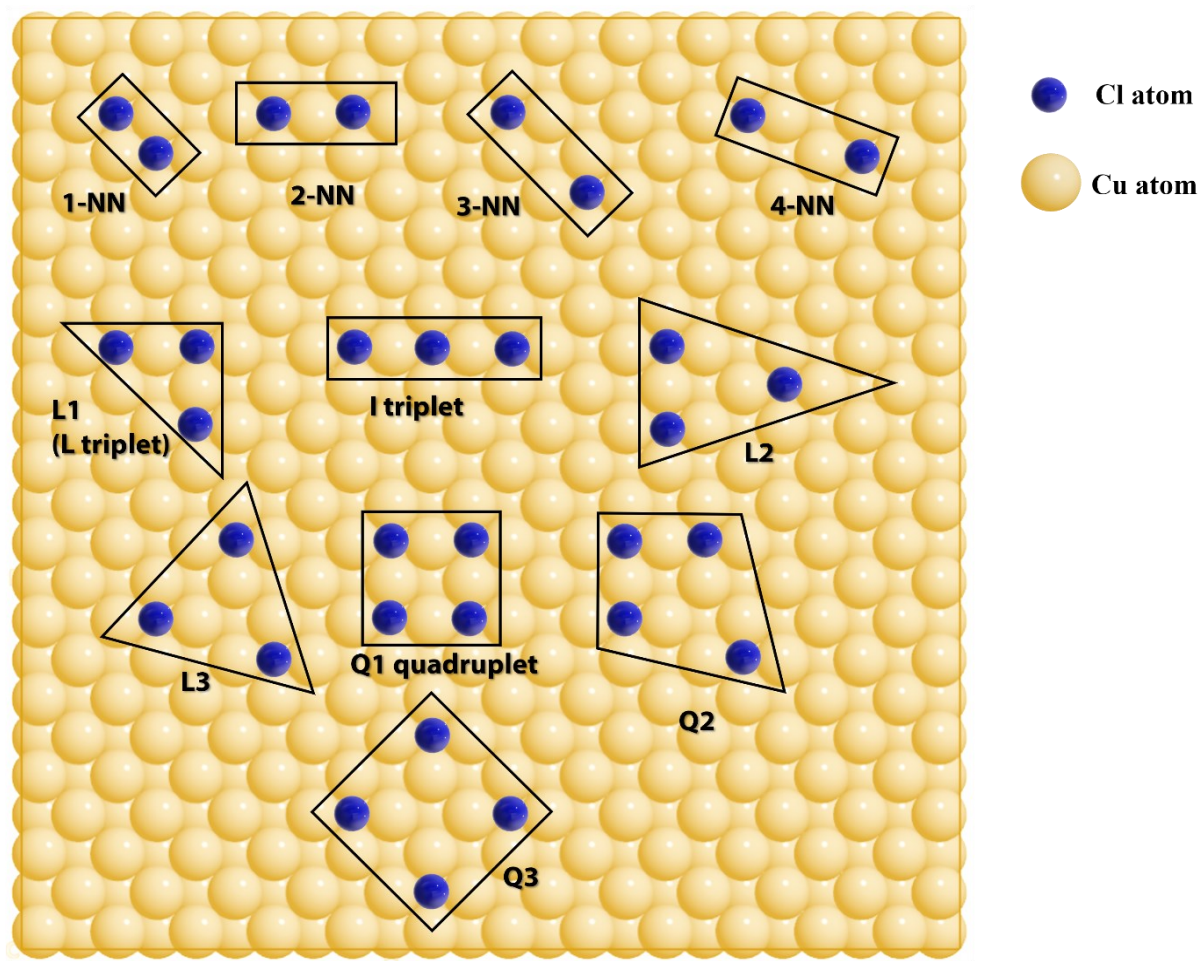

**Figure 2.** Different cluster types, such as pairs (1NN, 2NN, 3NN and 4NN), triplets (L1, L2, L3 and I) and quadruplets (Q1, Q2 and Q3) used in the CEM.

## *2.2 Cluster Expansion Model*

DFT calculations are computationally expensive. This limits the size of system that can be studied. Cu(100) surfaces typically considered in Monte Carlo simulations involve much larger lattice sizes, which allow for a number of Cl or Br adsorbed configurations. The energy $E_{nX/M}$ associated with a configuration involving a large lattice can be computed using Equation (1). The adsorption energy is evaluated with the help of a CEM. Thus, the CEM acts as a surrogate model to DFT. As mentioned in the Introduction section, the ECIs are obtained via numerical fitting exercise using a DFT database (Table S3 in Supporting Information). In literature, clusters typically included in CEM construction are singlet, pairs and triplets of different shapes [41,42,58–61]. For Cl and Br adsorption on $3 \times 3$ Cu (1 0 0) surface model used in our DFT calculations, 10 different clusters types are selected as shown in Figure 2. These include singlet, pairs (2NN, 3 NN, 4NN), triplets (three L-shaped and one I-shaped) and quadruplets (Q1, Q2 and Q3 cluster).

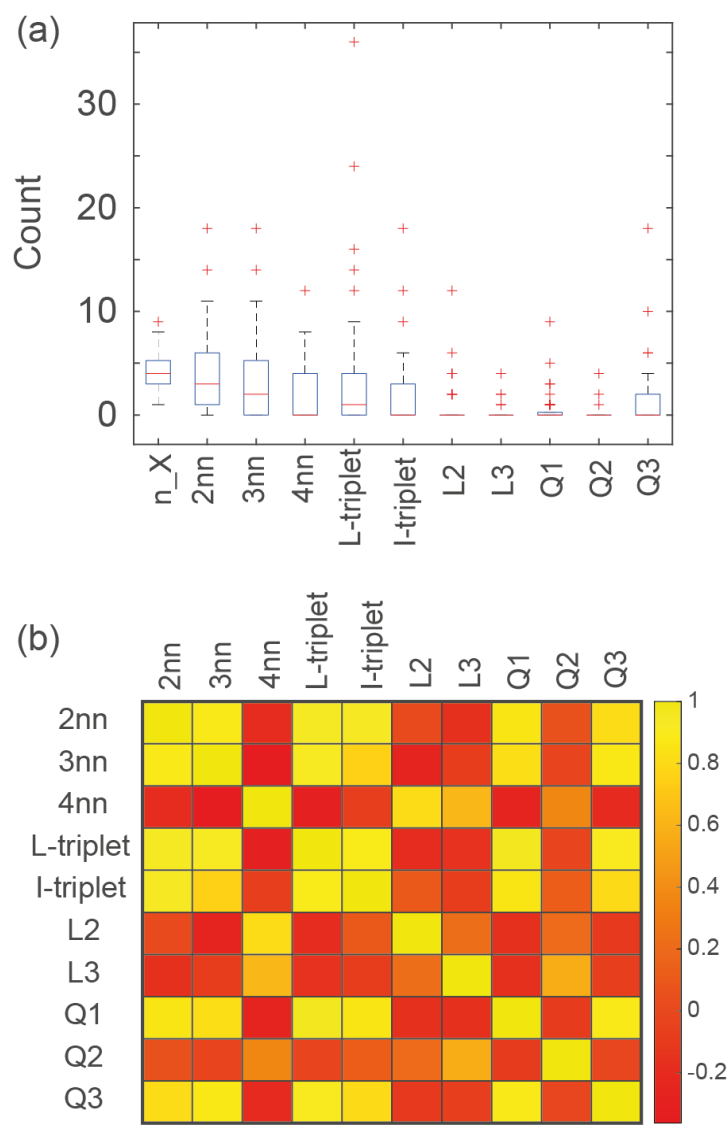

Figure 3. (a) Box plot for the cluster counts obtained with 42 different DFT configurations. (b) Corresponding correlation matrix for the cluster counts.

A number of halide configurations are generated by filling sites H1-H18 in Figure 1(b). Given the constraint that the 1NN position around an adsorbed Cl or Br cannot be occupied, the cluster count $\{n_c\}$ is evaluated for each configuration. Finally, 42 different configurations are identified that have unique cluster counts $\{n_c\}$. These configurations chosen for the DFT study. Figure 3a shows a box plot for the cluster counts. The number of X (=Cl/Br) atoms (denoted as $n\_X$ in the plot) varies between 1 and 9. The central mark in each box represents the median and the box edges are the $25^{th}$ and $75^{th}$ percentiles. Outliers are plotted using plus symbols. Figure 3b shows that a high degree of correlation can be found for certain clusters. For e.g., 2NN and 3NN pairs are strongly correlated to the L-triplet. On the other hand, 2NN and 4NN possess a negative correlation. As mentioned earlier, if these cluster counts are used directly in the regression process, collinearity will affect the outcome since the adsorbate-adsorbate interactions can be expressed in terms of the cluster counts in multiple ways.

We combine the cluster variables into principal components (see $\Lambda_p$ in Equation (5)) that are orthogonal to one another. Collinearity is eliminated in the process. Principal component regression (PCR) is a regression analysis technique based on PCA[51]. In this approach, the top few principal components ranked in terms of percentage variation explained are used as predictor variables for regression analysis instead of the original features.

*2.3 Grand Canonical Monte Carlo*

Grand Canonical Monte Carlo (GCMC) simulations are performed to understand the types of adsorbed phases with the resulting CEM. A 2D square lattice containing 50×50 site is used. Each site is either empty or occupied by a halide atom. A square lattice model is used because

(i) from DFT calculations, halides are found to preferentially bind at the hollow sites of Cu(100) surface and (ii) the halides are experimentally found to be arranged as a square lattice involving the hollow sites. Periodic boundary conditions are used.

The gas phase chemical potential $\mu_X$ and temperature $T = 77\ K$ is provided as an input. Each GCMC run consisted of 0.1 million trial moves, which was found to be adequate to reach equilibrium. Beyond this point the number of adsorbed halide atoms and energy is found to be nearly constant. Three different trial moves involving the halide atoms, namely, swap, insertion and deletion, are considered:

1. Swap: Randomly selected adsorbed halide atom is moved to a randomly selected vacant site with acceptance probability

$$p_{\text{acc},1} = \min(1, \exp(-\beta \Delta U)). \tag{6}$$

2. Insertion: A halide atom is inserted at a randomly selected vacant site with acceptance probability

$$p_{\text{acc},2} = \min\left(1, \frac{N_t - N_X}{N_X + 1} \exp(\beta \mu_X - \beta \Delta U)\right). \tag{7}$$

3. Deletion: A randomly selected halide atom is removed from the lattice with acceptance probability

$$p_{\text{acc},3} = \min\left(1, \frac{N_X}{N_t - N_X + 1} \exp-(\beta \mu_X + \beta \Delta U)\right). \tag{8}$$

Here, $N_t$ is the total number of lattice sites, $N_X$ is the number of filled sites, $\Delta U$ is the change in energy for the new configuration with respect to the old one when a move is attempted, $\beta = (k_B T)^{-1}$ and $k_B$ is the Boltzmann constant. GCMC requires the system energy ($E_{nX/M}$ in Equation (1)), which is calculated with the help of the CEM.

## 3. RESULTS AND DISCUSSION

## 3.1 Density Functional Theory

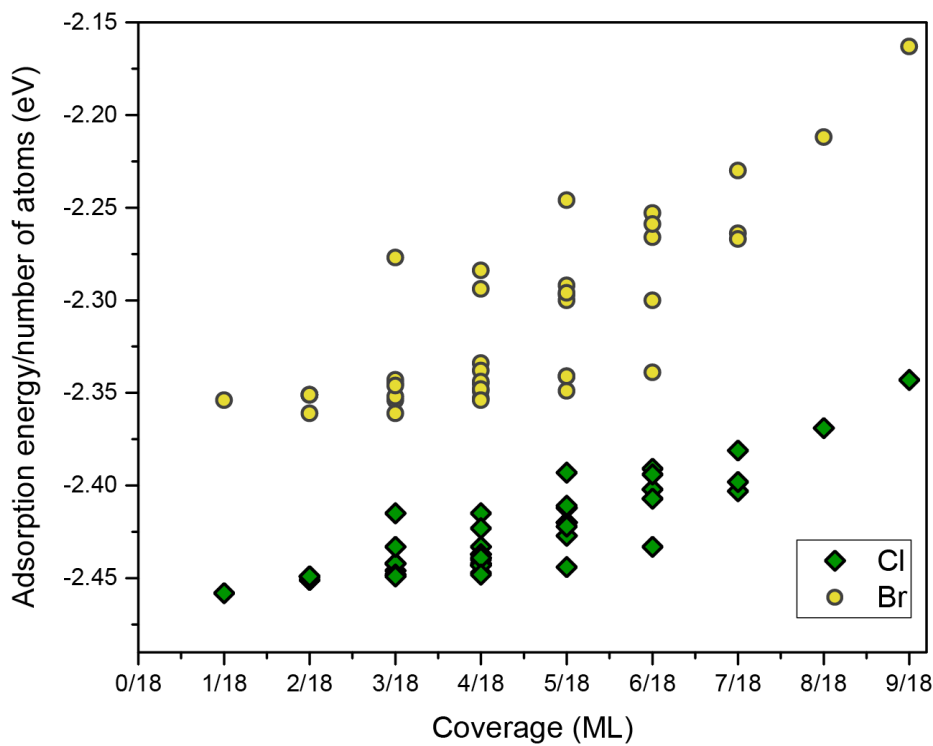

**Figure 4.** Adsorption energy for Cl and Br on Cu (100) calculated using DFT for different coverages.

Adsorption energy for Cl and Br at $\theta = \frac{1}{18}$ ML coverage is -2.458 and -2.354 eV, respectively, which sets the value of $E_{ads}^0$ in Equation (2). These values are consistent with ones reported in literature, namely, -2.43 eV and -2.31 eV, respectively [62] for the same coverage. Fully relaxed structures and energies of 42 configurations for Cl and similar number of configurations for Br are obtained using DFT with the halide coverages. The coverage effect on the adsorption energies of Cl and Br on Cu(100) are visible in Figure 4. As the $\theta$ increases, the adsorption energy per halide atom decreases indicating repulsive interactions. At the same time, from Figure 4 we see that the coverage itself is not a sufficient descriptor for the adsorption energy as there is a spread in the adsorption energy values for a specific coverage. Table S3 in Supporting Information lists the adsorption energy from DFT calculations. The occupation at the halide sites is mentioned at each of the hollow sites. Here 1 and 0 denote occupied and vacant sites, respectively.

*3.2 Cluster Expansion Model (CEM)*

Eigenvalues of the covariance matrix for cluster counts account for the multivariate variability. Table 1 shows the percentage variability explained after sorting the eigenvalues. The first four principal components (PCs) are able to explain more than 97% of the variability. Hence, these four are used for regression.

Table 1. Percentage of variance explained by the first n-principal components.

| Principal Component | Percentage variance explained |
|---|---|
| PC1 | 81.25 |
| PC2 | 11.40 |
| PC3 | 2.83 |
| PC4 | 1.79 |
| PC5 | 1.24 |
| PC6 | 0.81 |
| PC7 | 0.30 |
| PC8 | 0.23 |
| PC9 | 0.10 |
| PC10 | 0.04 |

Initially, 70% of the DFT dataset (29 randomly selected configurations out of 42) was used to train the CEM using 10 different kinds of cluster types as described in Section 2.2. The accuracy of the CEM was assessed by comparing the predicted adsorption energies with the DFT reference energies for remaining 30% of the configurations. The terms $e_b$ and $e_{singlet}$ correspond to the Langmuir adsorption model for non-interacting cases. The adsorbate-adsorbate interactions are captured by the remaining clusters. As discussed below, the predicted energies from the CEM are in very good agreement with the DFT values.

### 3.3.1 CEM for chlorine

To assess the model, we use both correlation coefficient ($R^2$ value) and a CV score. The CV score is defined as

$$Score = 1 - Normalized\ RMSE. \qquad (9)$$

The normalized RMSE is obtained by dividing the error for the many-body term $E_{ads-ads}$ by number of halide atoms. Figure 5a shows the model performance in terms of the number of PCs. The results change depending on how the initial dataset is split for training and testing purposes. For this particular sampling, the test set accuracy reaches a maximum at four PCs and then drops significantly thereafter. Since hold-out validation randomly splits the dataset, the training set sometimes may not have enough datapoints with a specific unique cluster. To address this uncertainty, we perform hold-out CV 1000 times. Each time we identify the number PCs required for the highest test score. In Figure 5b, the average score (score from the test set) of $i^{th}$ PC obtained from holdout cross validation is plotted. Figure 5b again shows that the average test score for 4 PCs are sufficient. The multivariate regression is performed with four PCs.

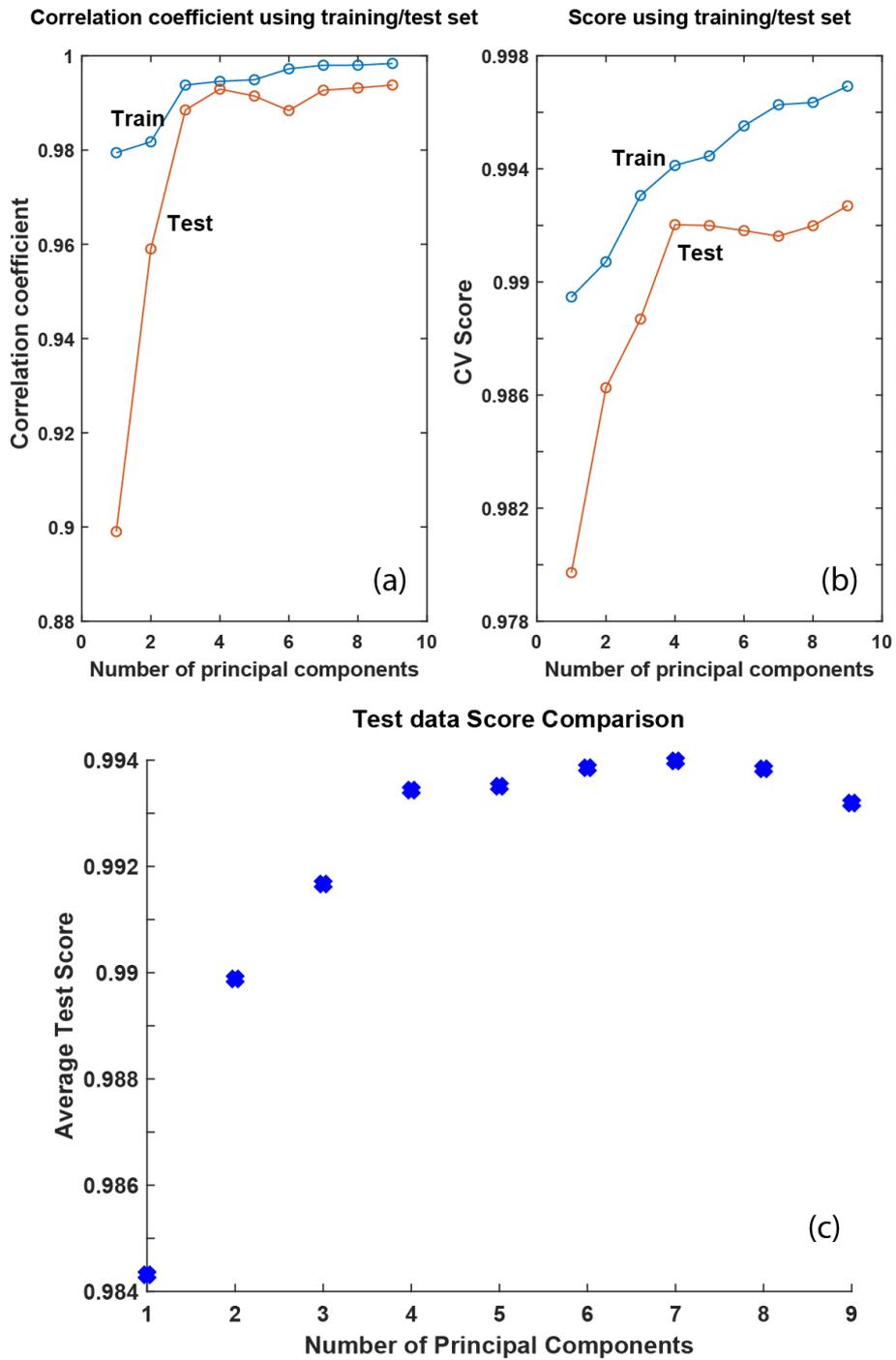

Figure 5. (a) Correlation coefficient ($R^2$ value) as a function of number of principal components included in the chlorine CEM. Results are shown for 70:30 split of full data into training and test sets. (b) Corresponding CV score (see Equation (9)). (c) Average performance of the trained CEM model.

Table 1. Effective cluster interactions for Cl/Cu(100). Standard deviation is mentioned. Split used for training:testing dataset for standard regression and principal component regression is shown in row 1. All values are in eV.

|  | Standard regression (70:30) | Principal Component Regression (70:30) | Principal Component Regression (20:80) |
|---|---|---|---|
| Adsorption energy/atom | −2.4579 | | |
| Bias | 0.0049 ± 0.0037 | 0.006 ± 0.0041 | 0.0108 ± 0.0165 |
| 2-NN | 0.012 ± 0.0029 | 0.0167 ± 0.0013 | 0.0189 ± 0.0062 |
| 3-NN | 0.0175 ± 0.0028 | 0.0018 ± 0.0017 | 0.0048 ± 0.0071 |
| 4-NN | 0.0065 ± 0.0016 | 0.0002 ± 0.0009 | 0.0012 ± 0.0083 |
| L-triplet | −0.0048 ± 0.005 | 0.0107 ± 0.001 | 0.0099 ± 0.0075 |
| I-triplet | 0.031 ± 0.0022 | 0.0163 ± 0.001 | 0.0159 ± 0.0051 |
| L2 | −0.0014 ± 0.0026 | 0.0091 ± 0.0008 | 0.0109 ± 0.0065 |
| L3 | −0.0154 ± 0.0039 | −0.0042 ± 0.0006 | −0.0039 ± 0.0031 |
| Q1 | 0.0224 ± 0.0129 | 0.0014 ± 0.0007 | 0.0004 ± 0.0042 |
| Q2 | 0.0101 ± 0.0043 | 0.0004 ± 0.0001 | 0.0006 ± 0.0008 |
| Q3 | −0.0069 ± 0.0032 | −0.0039 ± 0.0014 | −0.0046 ± 0.0079 |

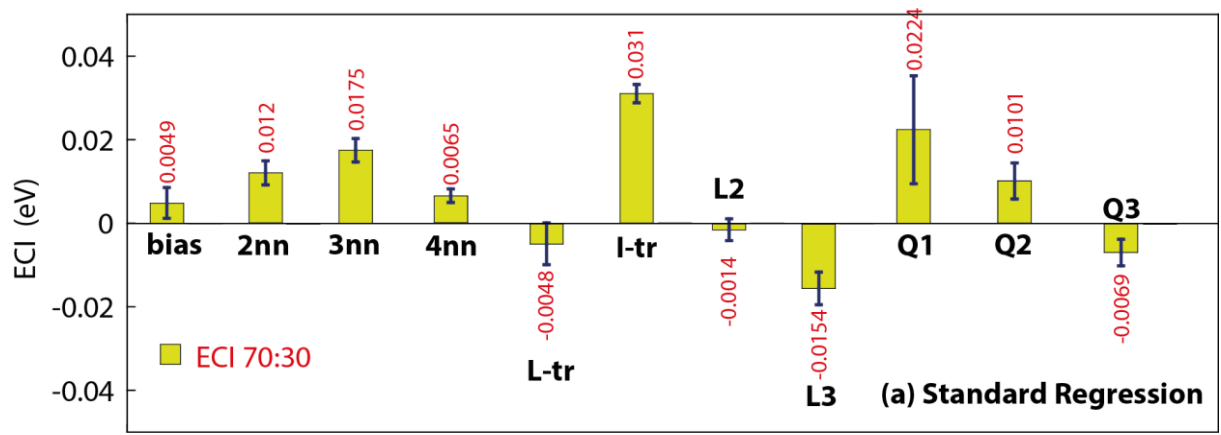

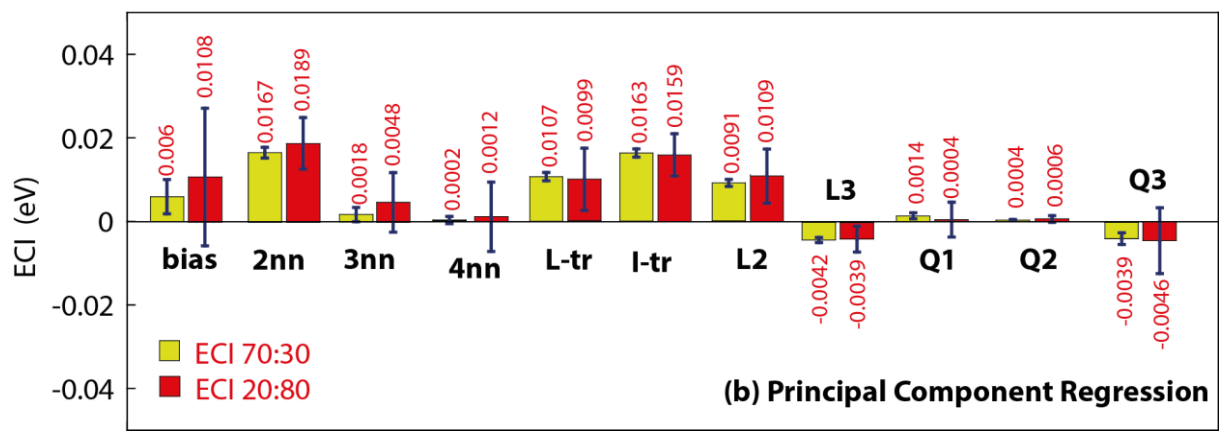

Figure 6. Comparison of effective cluster interactions (ECIs) obtained for Cl/Cu(100) using (a) standard and (b) principal component regression. Values are shown in red font and in Table 2. Two different training:test data partitioning are considered.

ECIs for both standard regression and (four PC based) PCR with 70% data used for training are shown in Table 2. Mean value and standard deviations for the interactions found over 1000 repeats are reported. Figure 6 shows the same results where the error bar corresponds to the standard deviation. The adsorption energy for a single Cl atom on Cu(100) is two orders of magnitude greater than the chlorine-chlorine interactions. This indicates that there is a strong tendency for the Cl atom to bind to the Cu surface. Since it is assumed that chloride ions do not occupy 1NN sites, a large positive ECI is used later in our GCMC calculations for the 1NN pair interaction, which promotes formation of the c(2×2) adlayer arrangement. Beyond 1NN, chlorine-chlorine interactions remain repulsive due to their strong electronegative character. Intuitively we expect that the pair ECI value should decrease as two adsorbed Cl atoms move apart. Results obtained from the standard regression (SR) model contradict this – ECI for 3NN is higher than 2NN. The PCR model captures the correct trend ($ECI_{2NN} > ECI_{3NN} > ECI_{4NN}$). Thus, PCR provides more physically realistic ECIs compared to standard regression. Additionally, the I-triplet and Q1 cluster interactions are strongly repulsive in the case of the SR model. To compensate for this, the L3 and Q3 cluster interactions are made attractive. In literature, this occurrence of large absolute ECIs is often incorrectly attributed to overfitting. In fact, such behavior is generally expected with standard regression due to collinearity. In contrast, triplet and quadruplet interactions from PCR are found to be weaker compared to the pair interactions. Q1 and Q2 interactions in the PCR-CEM are 10-100 times smaller than the ones in the standard regression model.

The low-dimensionality of the PCR-CEM suggests that smaller training dataset can be used. In Table 2 and Figure 6, results are also shown for the case where only 8 DFT energies are used for training the CEM. In this case, the training:test dataset is 20:80. Such a training dataset cannot be used with standard regression as the number of unknown coefficients exceeds the number of datapoints. However, the CEM with 4 PCs can be trained. In Figure 6, we observe that the ECIs obtained from such an exercise are nearly same as the ones obtained from a 70:30 split. The standard deviation for the ECIs has slightly increased. Table 2 lists the values of the ten ECIs and the bias obtained from the 8 DFT training energies.

### 3.3.2 CEM for bromine

A similar methodology is used to train CEM for Br-Br interactions on Cu(100) surface. From previous section, it was determined that 4 PCs are sufficient. Figure 7a shows that the correlation coefficient ($R^2$ value) obtained using 70% of the DFT data for model training exceeds 0.99 for both training and testing when 4 PCs are included. Figure 7b shows that the CV score obtained is also reasonable. Figure 7c shows the average test score from multiple runs exceeds 0.987. Therefore, similar to the case of chlorine we proceed with four principal components.

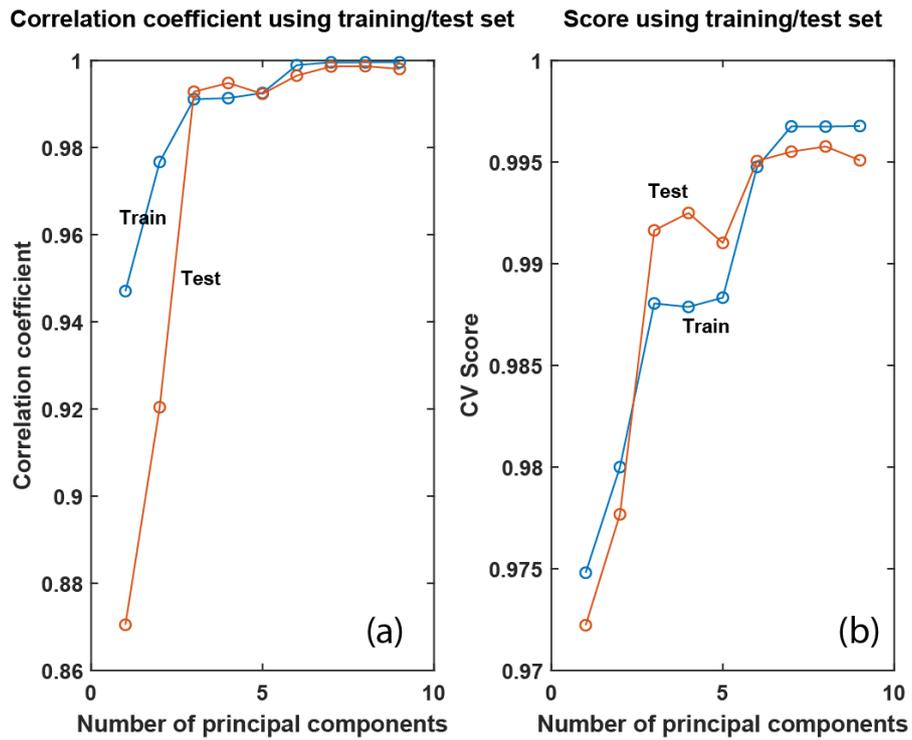
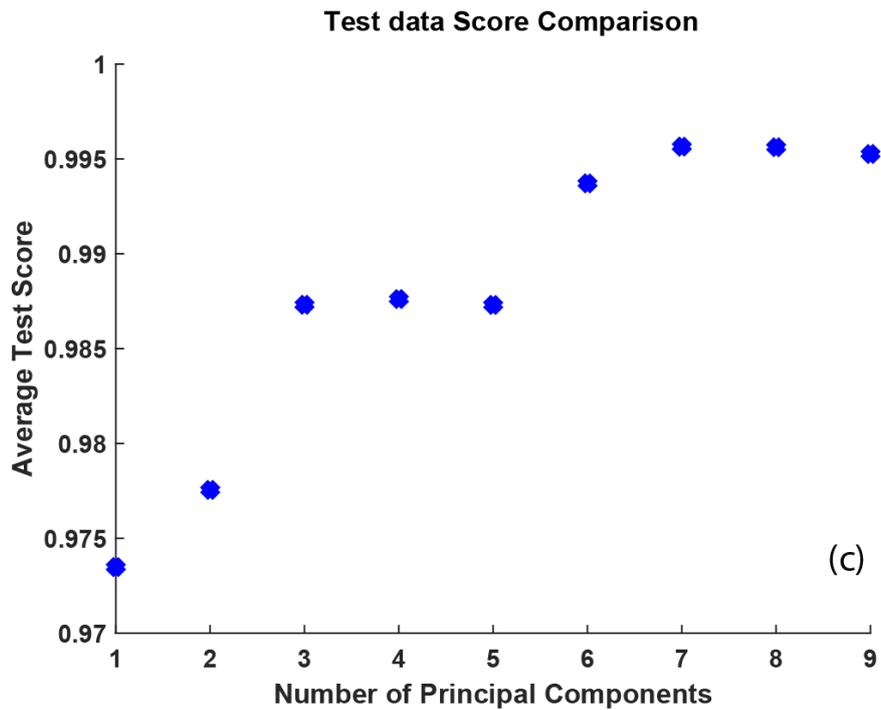

Figure 7. (a) Correlation coefficient ($R^2$ value) as a function of number of principal components included in the bromine CEM. Results are shown for 70:30 splitting of full data into training and test sets. (b) Corresponding CV score (see Equation (9)). (c) Average performance of the trained CEM model.

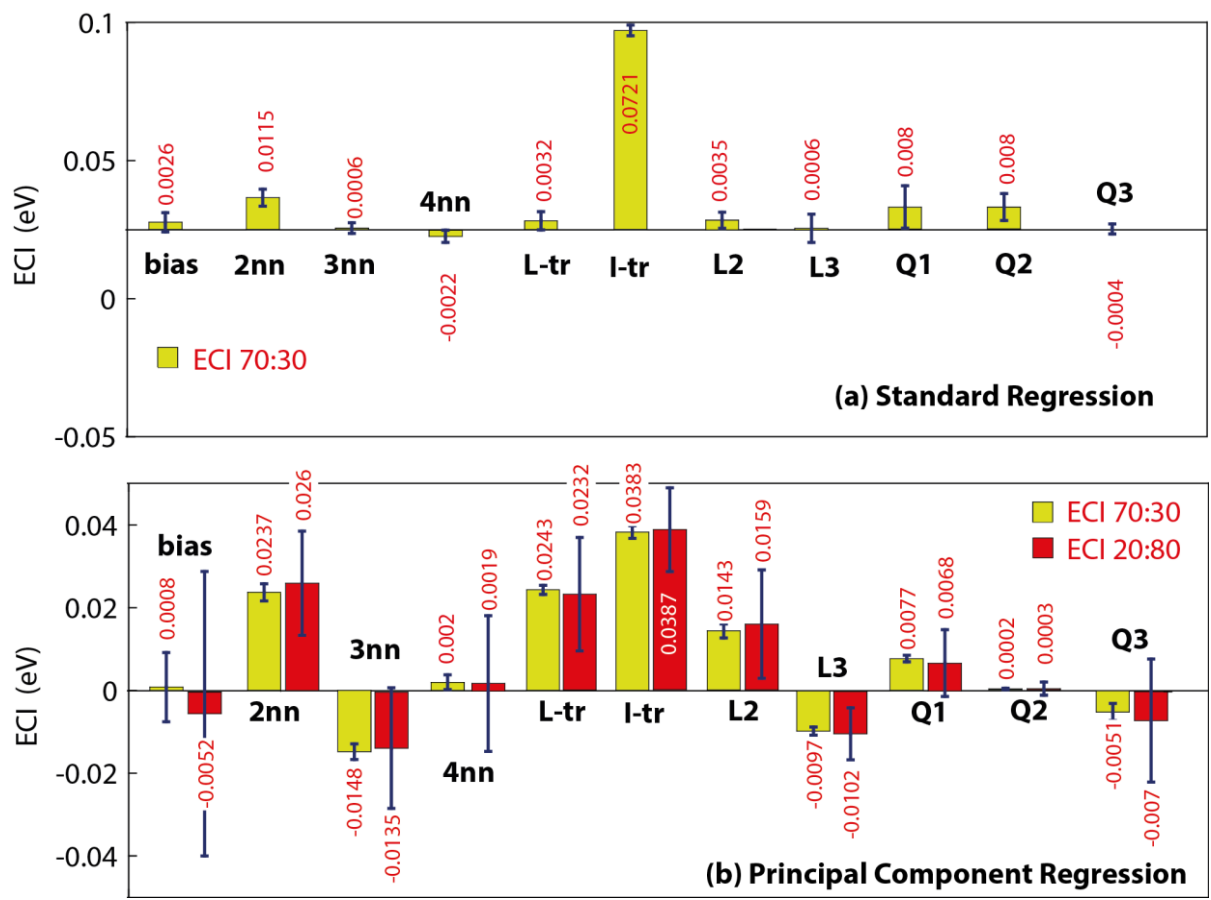

Figure 8. Comparison of effective cluster interactions (ECIs) obtained for Br/Cu(100) using (a) standard and (b) principal component regression. Values are shown in red font and in Table 3. Two different training:test data partitioning are considered.

A comparison of ECIs from standard regression and PCR for Br/Cu(100) is shown in Figure 8 and Table 3. A single Br atom adsorbs slightly weakly compared to Cl. Figure 4 shows that that the presence of neighbor Br atoms results in steeper increase in the adsorption energy compared to Cl. This suggests that the Br-Br interactions should be more repulsive than in the case of Cl. In the case of standard regression, ECIs are generally found to be repulsive. An odd observation is that I-triplet (0.072 eV) is found to be exceptionally repulsive than others. Although ECI from PCR is also repulsive in nature, such large values of ECI are not observed. L-triplet and I-triplet clusters have interaction strengths of 0.02-0.03 eV. Recall that in case of Cl it was similarly observed that the ECIs are overestimated by standard regression. Interestingly, 3NN pairs, L3 and Q3 interactions are found to be attractive in PCR. Similar values of ECI strengths are obtained when the ratio of the training to test data is 20:80. The standard deviation in the ECIs for the 20:80 case is large when compared to the 70:30 ratio.

Table 2. Effective cluster interactions for Br/Cu(100). Split used for training:testing dataset for standard regression and principal component regression is shown in row 1. All values are in eV.

| | Standard regression (70:30) | Principal Component Regression (70:30) | Principal Component Regression (20:80) |
|---|---|---|---|
| Adsorption energy/atom | 2.354 | | |
| Bias | $0.0026 \pm 0.0035$ | $0.0008 \pm 0.0084$ | $-0.0052 \pm 0.0344$ |
| 2-NN | $0.0115 \pm 0.0031$ | $0.0237 \pm 0.0021$ | $0.026 \pm 0.0126$ |
| 3-NN | $0.0006 \pm 0.0019$ | $-0.0148 \pm 0.0019$ | $-0.0135 \pm 0.0146$ |
| 4-NN | $-0.0022 \pm 0.0022$ | $0.002 \pm 0.0018$ | $0.0019 \pm 0.0164$ |
| L-triplet | $0.0032 \pm 0.0033$ | $0.0243 \pm 0.0011$ | $0.0232 \pm 0.0137$ |
| I-triplet | $0.0721 \pm 0.0019$ | $0.0383 \pm 0.0015$ | $0.0387 \pm 0.0101$ |
| L2 | $0.0035 \pm 0.0029$ | $0.0143 \pm 0.0017$ | $0.0159 \pm 0.0131$ |
| L3 | $0.0006 \pm 0.0051$ | $-0.0097 \pm 0.001$ | $-0.0102 \pm 0.0063$ |
| Q1 | $0.008 \pm 0.0077$ | $0.0077 \pm 0.0008$ | $0.0068 \pm 0.0081$ |
| Q2 | $0.008 \pm 0.0049$ | $0.0002 \pm 0.0002$ | $0.0003 \pm 0.0016$ |
| Q3 | $-0.0004 \pm 0.0018$ | $-0.0051 \pm 0.0021$ | $-0.007 \pm 0.0149$ |

*3.3 Grand Canonical Monte Carlo simulations*

### 3.4.1 Chlorine adsorption

GCMC loading and unloading simulations are performed with a 2D square lattice to obtain the equilibrium halide adlayer configuration on copper. The terms loading and unloading refer to the initial coverage for the simulation being zero and 0.5 ML, respectively. In the latter case, the initial configuration corresponds to a $c(2 \times 2)$ structure. The chemical potential is kept at a fixed value in each GCMC calculation. The chemical potential of halide is varied between -2.7 to -1.5 eV, with a step of 0.02 eV. Two different CEMs are used in GCMC. These are denoted as CEM-E referring to extensive training dataset (with training:test data being 70:30) and CEM-L referring to limited training dataset (20:80). These models were discussed in Figure 6b. The main purpose is to show that (i) it is important to include the absorbate-adsorbate interactions and (ii) results obtained with CEM-L are comparable to ones from CEM-E.

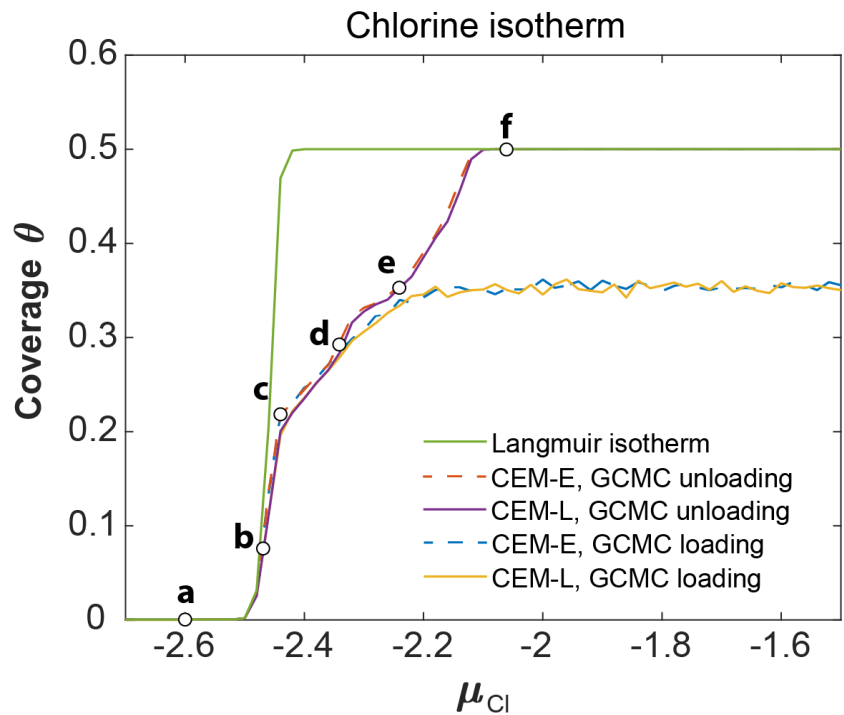

Figure 9. Adsorption isotherm for Cl/Cu(100). Chlorine adlayer arrangement in unloading simulations is shown in Figure 10 at different values of $\mu_{Cl}$, namely, (a) $-2.6$, (b) $-2.48$, (c) $-2.44$, (d) $-2.34$, (e) $-2.24$ and (f) $-2.06\ eV$.

Figure 9 shows the Cl/Cu(100) adsorption isotherm. Any coverage in the range of 0-0.5 is attainable. Six different coverages are labelled as points (a)-(f). The corresponding Cl configuration is shown in Figure 10. If we consider only the binding energy term $E^0_{ads}$ and exclude the absorbate-adsorbate interactions, GCMC simulation yields the Langmuir isotherm (green line in Figure 9). The coverage obtained from CEM-E and CEM-L matches with the Langmuir isotherm for coverage less than 0.1 ML. This implies that the absorbate-adsorbate interactions are non-existent (points (a) and (b)) in Figure 9). The adsorbed Cl atoms are randomly distributed and far apart (Figure 10b). Absorbate-adsorbate interactions become relevant beyond 0.2 ML coverage (point (c)). The repulsive interaction causes the absorption isotherm to shift to the right with respect to the Langmuir isotherm, i.e., higher chemical potential is required to achieve a coverage of 0.2 ML. Figure 10c shows the formation of a local ordered overlayer structure ($M = \begin{bmatrix} 2 & 1 \\ 0 & 2 \end{bmatrix}$ in the matrix notation). At a coverage of 0.3 ML, a $p(2 \times 2)$ overlayer structure is observed (Figure 10d). At 0.35 ML, yet another overlayer structure is formed, namely, $M = \begin{bmatrix} 2 & 0 \\ -1 & 1 \end{bmatrix}$ (Figure 10e or point (e)) and $c(2 \times 2)$. The overlayer structures from GCMC contain defects, span short spatial distances and are not contiguous. These overlayer regions are also associated with kinks in the adsorption isotherm. A perfect $c(2 \times 2)$ overlayer structure is obtained at 0.5 ML (Figure 10f). Similar $c(2 \times 2)$ overlayer structures have been reported in STM images of experimental surfaces [26,30,31].

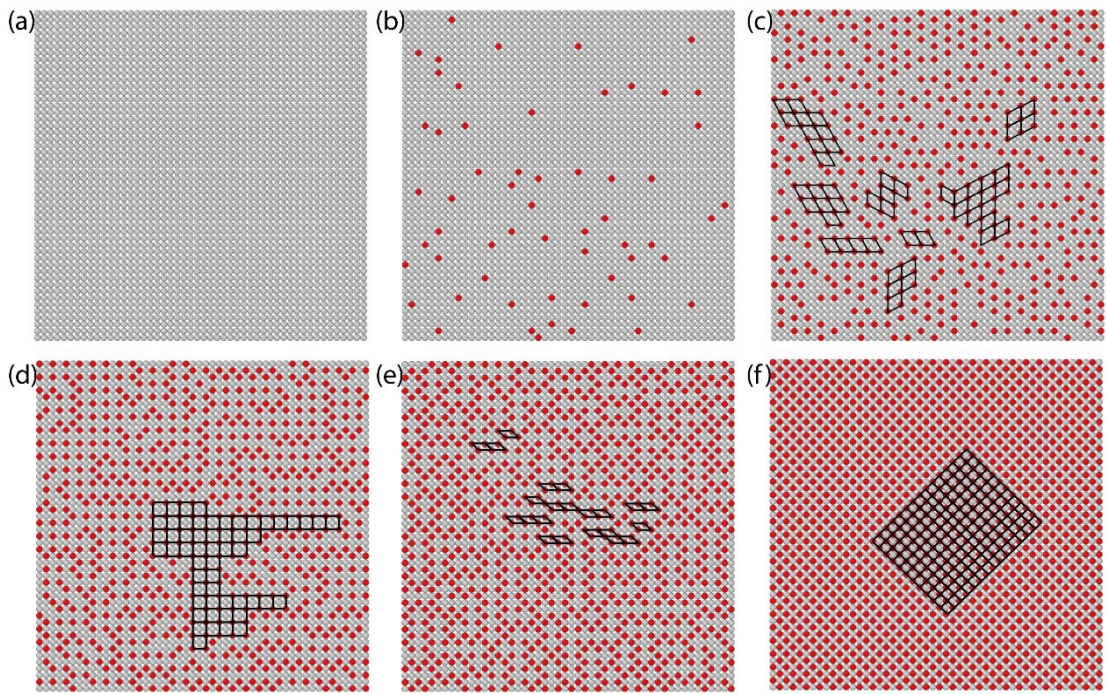

Figure 10. Cl adlayer configurations from unloading GCMC simulations. Panels refer to the points in Figure 9. Grey spheres are hollow sites, red denotes Cl atoms. Overlayer structures are shown in lines.

An important observation is that in spite of the small differences in the ECI values, both CEM-L and CEM-E provide the same (loading and unloading) absorption isotherm. This suggests that the thermodynamic behavior is not severely affected by the uncertainty in the model parameters (see error bars in Figure 6b). The loading and unloading curves begin to diverge at point (d). Coverages beyond 0.354 ML are not encountered in the loading simulations. The reason for this is entropic in nature. At 0.35 ML coverage, the overlayer structures contain a high concentration of defects. These defects are associated with large entropy and a large insertion energy for a Cl atom. Although a perfect $c(2 \times 2)$ structure is energetically favorable due to the high binding energy of Cl on Cu, configurations that allow for long-ranged ordering are not sampled in the GCMC simulation. A perfect $c(2 \times 2)$ structure is obtained at high chemical potentials only with unloading simulations.

### 3.4.2 Bromine adsorption

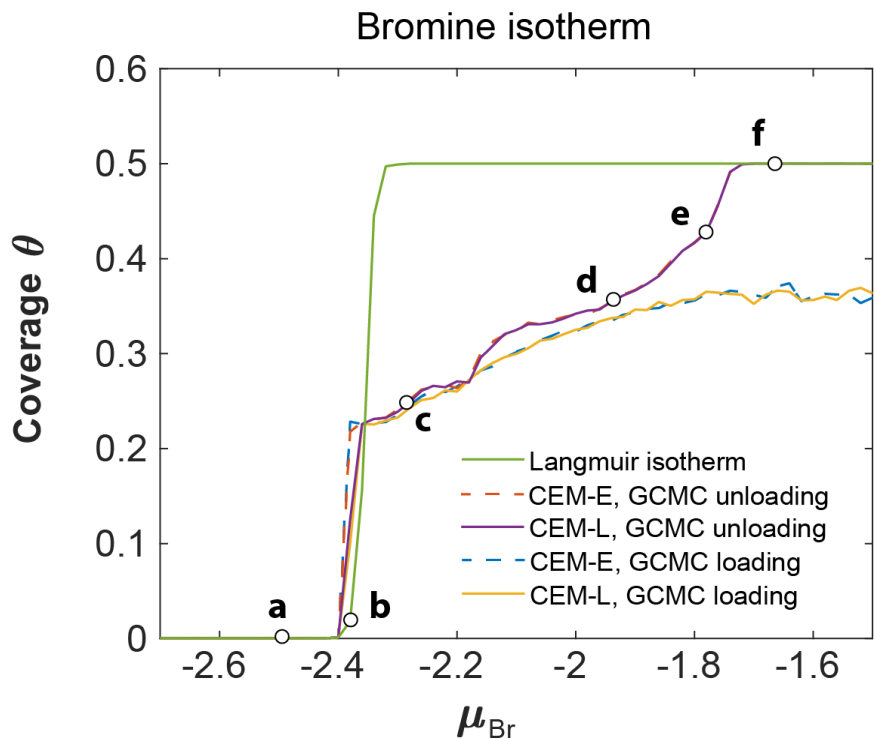

Figure 11. Adsorption isotherm for Br/Cu(100). Bromine adlayer arrangement in unloading simulations is shown in Figure 12 at different values of $\mu_{Br}$, namely, (a) $-2.5$, (b) $-2.38$, (c) $-2.28$, (d) $-1.94$, (e) $-1.78$ and (f) $-1.66\ eV$.

Adsorption of bromine on copper exhibits a behavior similar to Cl. Because of the weaker binding energy $E_{ads}^0$, the Langmuir isotherm is shifted to the right compared to the Cl isotherm. Because of the stronger repulsive interactions, overall, the deviation from the Langmuir isotherm (green line in Figure 11) is even more pronounced for Br. An important difference is that Br atoms are able to cluster even at low coverages (see Figure 12b). A local ordered overlayer structure ($\boldsymbol{M} = \begin{bmatrix} 2 & 1 \\ 0 & 2 \end{bmatrix}$ in the matrix notation) is observed at 0.13 ML coverage. Subsequently, three partial overlayer structures are seen, namely, $p(2 \times 2)$ at 0.25 ML, $\boldsymbol{M} = \begin{bmatrix} 2 & 0 \\ -1 & 1 \end{bmatrix}$ at 0.37 ML and $c(2 \times 2)$ at 0.42 ML. A complete $c(2 \times 2)$ overlayer structure is seen at 0.5 ML coverage. Similar to Cl, the loading and unloading curves begin to diverge. In case of Br this happens sooner once the coverages are greater than 0.26 ML.

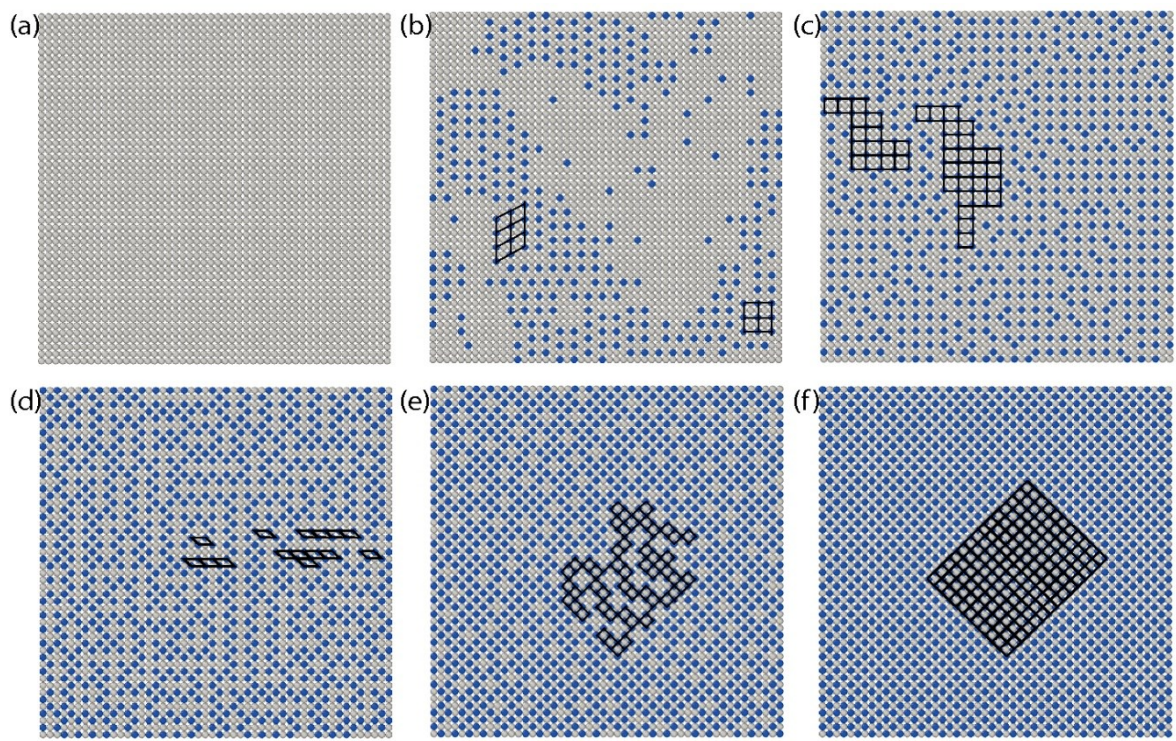

Figure 12. Br adlayer configurations from unloading GCMC simulations. Panels refer to the points in Figure 11. Grey spheres are hollow sites, blue denote Br atoms. Overlayer structures are shown in lines.

## 4. Conclusion

Adsorbate-adsorbate interactions involving species such as Cl and Br, bound to a metal surface like copper, can be complicated, many-body and repulsive in nature. These interactions influence the adsorption behavior. For example, the adsorption isotherm obtained is different from the Langmuir isotherm and ordered adlayer structures are formed. The cluster expansion model (CEM) framework is commonly used to capture adsorbate-adsorbate interactions. The main accomplishment of this paper is the use of principal component regression (PCR) technique to build a low-dimensional, reduced collinearity CEM for adsorption of halides (Cl or Br) on Cu(100) surface. The CEM is low-dimensional since only a small number of principal components are employed. Like in conventional CEM training, sites within a cutoff radius are considered. However, our approach supports the unrestricted incorporation of all cluster types formed by these sites, to the extent that it is practical for the user. The results obtained highlights that not only it is possible to construct high-fidelity models with limited DFT data, but the number of data points required for training can in fact be fewer than the actual number of ECIs in the model. Such a capability is particularly useful when multiple adsorbed species are simultaneously present at the surface. One can include a large variety of atom clusters in the CEM while retaining only the important principal components.

To summarize, the novel utilization of PCR replacing simple multivariate regression offers the following advantages:

1) User can incorporate in the CEM as many cluster types, pairs, triplets, quadruplets and so on, as required based on chemical intuition or without, while exploiting the ability of the PCR technique to eliminate the correlations during the regression procedure. By retaining all cluster types, information loss is prevented.

2) Compared to standard regression, physically meaningful values are obtained with PCR for the effective cluster interactions, and

3) The method is expected to have important application for more complex situations, such as co-adsorption, where one can also quickly estimate the number of DFT simulations required using PCA. Once more, it should be possible to train the CEM with limited amount of DFT data.

To demonstrate the application of these ideas to *ab initio* thermodynamics modelling, we have employed the CEM in grand canonical Monte Carlo (GCMC) simulations to find the adlayer halide overlayer structures on Cu(100). The simulations successfully reproduce experimental observed adsorbed configurations validating the CEM.

## 5. Acknowledgements

This work benefitted from the helpful discussions with Prof. Sharad Bhartiya and Prof. Sujit S. Jogwar. AC acknowledges support from Science and Engineering Research Board, Grant Nos. CRG/2022/008058 and National Supercomputing Mission DST/NSM/R&D_HPC_Applications/2021/02.

**Supporting Information**

The Supporting Information is available free of charge on the ACS Publications website at DOI: .

Tables providing binding energy for different halide on copper configurations.

**TOC Image**

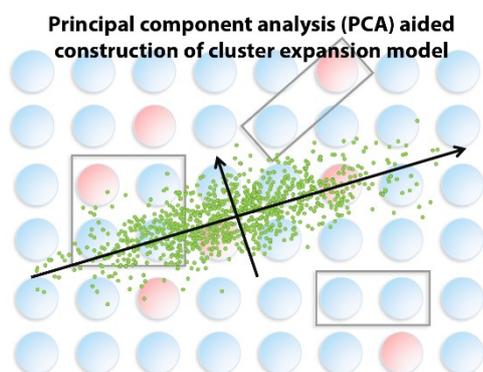